\documentstyle[prl,aps]{revtex}
\input{epsf}
\draft
\begin{document}
\twocolumn[\hsize\textwidth\columnwidth\hsize\csname@twocolumnfalse\endcsname
\title{ Helicoidal instability of a scroll vortex 
  in three-dimensional reaction-diffusion systems}
\author{Igor Aranson$^{1}$ and  Igor Mitkov $^{2}$}
\address{
$^1$~Argonne National Laboratory,
9700 South  Cass Avenue, Argonne, IL 60439 \\
$^2$~Center for Nonlinear Studies and
Computational Science Methods Group\\
Los Alamos  National Laboratory,
Los Alamos, NM 87545  
}
\date{\today}
\maketitle

\begin{abstract}
We  study the dynamics of scroll vortices in excitable 
reaction-diffusion systems
analytically and numerically.
We demonstrate  that  
intrinsic three-dimensional instability of a straight scroll
leads to the formation of helicoidal structures. 
This behavior  originates from the competition between the scroll
curvature and unstable core dynamics. 
We show that the obtained instability persists even beyond 
the meander core instability of two-dimensional spiral wave.   
\end{abstract}

\narrowtext
\vskip1pc]

Spiral waves arising in  diverse  physical, chemical,
and biological systems are now one of the paradigms of
non-equilibrium dynamical  phenomena \cite{ch}. Examples include the Belousov-
Zhabotinsky (BZ) reaction~\cite{Skinner} and catalytic oxidation
of CO on Pt substrates~\cite{Jakubith},
concentration waves in colonies of aggregating
amoebae~\cite{amoeba}, waves in cardiac tissue~\cite{Allessie},
and many others. These seemingly
unrelated phenomena
share a common feature: they often allow for a description within
the framework of two-component
reaction-diffusion type  systems, for which spiral solutions are generic.

The three-dimensional analog of a spiral wave is a scroll vortex, which can be 
represented by translating the spiral wave along the third direction. 
Thus, the point singularity of two-dimensional spiral wave (tip) develops 
into a line singularity (vortex filament). The dynamics of the vortex 
filaments in reaction diffusion systems has attracted a great deal of
attention~\cite{biktashev,panfilov} in connection with sudden heart
fibrillation, where it is believed that
scroll and ring vortices play a crucial role~\cite{panfilov1,winfree}.
Extensive numerical
simulations~\cite{biktashev,panfilov,panfilov1} and recent experiments
on gel-immobilized  BZ reaction~\cite{winfree2,pertsov1} 
show that in a wide range of parameters the scrolls  are unstable and 
may assume  helicoidal or even more complicated dynamic 
configurations.  
However, the theoretical analysis performed first by 
J.~Keener, predicts an ultimate collapse of vortex rings and stability of
straight vortex filaments~\cite{keener}.
This conclusion was drawn on the basis of 
multiscale analysis, which shows that a vortex ring is similar to 
an elastic line  with a positive line tension. 
The persistence of nontrivial vortex configurations and turbulence 
in reaction-diffusion systems was attributed 
to a negative line tension of the filament~\cite{biktashev}.

In this Letter we demonstrate,
on the basis of numerical and analytical calculations,
that the formation of helicoidal vortices
can be related to the
intrinsic three-dimensional
instability of a straight scroll, caused by
a nontrivial response of the filament core to a bending of the
filament.
We show that the limit of this instability, resulting in
the formation of spontaneous helicoidal vortices,
goes beyond the corresponding two-dimensional core meander instability.

The dynamics of a scroll vortex can be consistently described by 
the two-component reaction-diffusion system
\begin{eqnarray}
&&\partial_t u=\epsilon \nabla ^2u+\frac{f(u,v)}\epsilon \;,
\label{eq1}\\
&&\partial_t v=\delta\epsilon \nabla ^2v+g(u,v) \;,
\label{eq2}
\end{eqnarray}
where $\epsilon$ is a small positive parameter, and
$\delta=D_v/D_u$ is the ratio of diffusion coefficients of
the variables $v$ and $u$. The functions $f(u,v)$ and $g(u,v)$
are chosen so as to make Eqs. (\ref{eq1})--(\ref{eq2})
{\it excitable}~\cite{ch}. In a wide range of parameters 
Eqs. (\ref{eq1})--(\ref{eq2}) have a spiral wave solution in
two dimensions and a scroll vortex  in three dimensions. 

Let us derive the equation of motion for the core of a weakly-curved scroll
vortex subject to the meandering instability~\cite{Skinner,winfree1}.
In two dimensions this instability was studied both via numerical simulations
of the system (\ref{eq1})--(\ref{eq2}), and using
the numerical solution of the linearized problem \cite{barkley1,barkley3}.
Recently it was proven that the spiral interface undergoes a core-meander
instability via a supercritical Hopf bifurcation, as the diffusion
coefficient of the slow field $\delta$ decreases~\cite{MAK}.
The mechanism of the meandering instability in a certain parameter limit
has been elaborated on in a recent work~\cite{karma}.

The core meander was  described by five
equations for the frequency, coordinates, and the velocity
of the spiral tip \cite{barkley3}. 
These  equations can be considerably simplified if, instead
of the spiral tip, one considers the instant center of spiral rotation.
At the threshold of instability one obtains a
single complex  Landau-type equation for 
the ``complex'' velocity of the rotation center   
$\hat{C} = c_x + i c_y$, where $c_{x,y}$ are the components of the velocity: 
\begin{equation}
\partial_t \hat{C} = \alpha \hat{C} - \beta |\hat{C}|^2 \hat{C} \;.
\label{eq3}
\end{equation}
Here $\alpha$ and $\beta$ are complex coefficients,
$\alpha = \alpha_1 + i\alpha_2\,,\, \beta = \beta_1 + i\beta_2\,$,
that have to be determined numerically.
For $\alpha_1 < 0$ the symmetry center is stable,
and the spiral makes a pure rotation. When $\alpha_1$ becomes positive, 
the fixed point solution of Eq.~(\ref{eq3}) loses stability,
and the rotation center itself performs a circular motion,
which implies the meandering (composite rotation)
of the spiral tip \cite{comm}. As follows from Eq.~(\ref{eq3}),
the absolute value of the velocity of the rotation center,
in the saturated meandering regime,
is $C_0 = \sqrt{\alpha_1/|\beta_1|}$,
and the corresponding rotation frequency is 
$\omega_0 = \alpha_2 - \alpha_1\beta_2/\beta_1\,$.

Consider now a three-dimensional weakly curved scroll 
(the curvature $\kappa$ in the third dimension is much
smaller then the local curvature of the spiral front). 
In this limit one expands the Laplacian in
equation~(\ref{eq1}) as follows,
$\nabla^2 \approx \nabla^2_{2D} - \kappa {\bf N}
\cdot \nabla$,
where $\nabla^2_{2D}$ is the Laplacian in the cross-section, and
${\bf N}$ is the unit vector pointing toward the center of the
filament curvature. Therefore, the  curvature
in the third dimension plays the role of an advective field 
directed along ${\bf N}$ and causing
the drift of the filament \cite{MAK1}. 
To lowest  order, the curvature $\kappa$
enters into the equation of motion (\ref{eq3}) linearly: 
\begin{equation} 
 \partial_t\hat{C} = \alpha\hat{C} -\alpha\gamma\kappa + ...
\label{eq5}
\end{equation} 
where $ \gamma = \gamma_1 + i\gamma_2$ is a (complex) constant which can
be determined numerically from two-dimensional simulations \cite{MAK1} 
or analytically in a large core limit \cite{karma}.

Equation (\ref{eq5}) readily implies the short-wavelength
instability of a straight filament. Indeed, for an almost straight filament
parallel to the $z$-axis, the curvature vector $\kappa {\bf N}$ is simply 
$(x_{zz}, y_{zz})$.  
Using $\hat{C} = \partial_t\hat{x}\,$, where $\hat x = x+i y $,
we obtain from Eq.~(\ref{eq5}), for a periodic  perturbation
of the filament $\hat{x}(z) \sim \exp[\lambda(k)t + i k z]\,$,
that $\lambda^2 = \alpha(\lambda + \gamma k^2)\,$.
The latter equation has the following unstable solution
\begin{equation}
\lambda(k) = \frac{\alpha}{2} + \sqrt{\frac{\alpha^2}{4}
+\alpha\gamma k^2}\;.
\label{eq6}
\end{equation}
Eq.~(\ref{eq6}) represents the eigenvalue of the linearized
problem (\ref{eq5}) and
is valid only for slightly curved filaments, {\it i.e.} for
small wavenumbers $k\,$.
If this near-threshold instability saturates,
the saturated structure corresponds to the most unstable mode,
which implies the formation of stable helicoidal vortices.

By analogy with the vortex filaments in the CGLE~\cite{aranson},
we can expect that, for $k > 0\,$, the growth rate of the obtained
three-dimensional instability, $Re[\lambda(k)]$, substantially
exceeds that of the two-dimensional meandering instability,
$Re[\lambda(0)] = Re[\alpha]$. In that case the filament curvature
plays a destabilizing role in the dynamics of the filament.
To analyze the stability, one can find the coefficients
$\alpha$ and $\gamma$ from the dynamics of the two-dimensional
system~(\ref{eq1})--(\ref{eq2}), and then substitute them into
Eq.~(\ref{eq6}).

To determine $\alpha$ and $\gamma$ numerically, we have simulated
Eqs.~(\ref{eq1})--(\ref{eq2}) in two dimensions using
the EZ-spiral code of D.~Barkley~\cite{barkley2}.
We have chosen Barkley's model with the functions
$f(u,v)=u(u-1)[u-u_{th}(v)]$ and $g(u,v)=u-v$, where $u_{th}(v)=(v+b)/a$.
To determine numerically $\alpha, \beta$ and $\gamma$,
one should start with an unstable rigidly rotating spiral, 
which is not available in numerical simulations.
To overcome this difficulty,
we used the following approach. We started with a spiral
with already developed meander. Then we applied a
{\em localized control technique}, developed in Ref.~\cite{ALT},
to turn the spiral
motion to a pure rotation around its symmetry center.
Following Ref.~\cite{ALT}, we applied a {\em pinning source}
to equation~(\ref{eq2}), in the form of a localized inhibiting
term $-\mu\, h(r-r_0)\,$, where $r_0$ is the coordinate of the domain center.
Here $\mu$ plays the role of a {\em control parameter} and is
governed by the equation
\begin{equation}
\partial_t \mu = -a_1\mu + b_1 [v(r_0) - v_0]\;,
\label{eq7}
\end{equation}
where $v_0$ is the expected value of the slow field at the
spiral center, and $a_1$, $b_1$ are coefficients
that should be properly chosen. We have elaborated the controlling
technique by allowing $v_0$ to vary adiabatically
(slowly compared to $\mu$) as follows
\begin{equation}
\partial_t v_0 = a_2\mu + b_2 [v(r_0) - v_0]\;,
\label{eq8}
\end{equation}
with another pair of coefficients $a_2$, $b_2$.
In this way, we have obtained a purely rotating spiral
with the parameter values corresponding to a meandering
regime. The final value of the control parameter was as small as
$\mu=10^{-4}\,$, which made the source term added to equation~(\ref{eq2})
a small perturbation. After ``switching off'' the control
(setting $\mu=0$), we allow the linear meander instability to develop.

\begin{figure}[h]
\hspace{0.2cm}
\rightline{ \epsfxsize = 9.0cm \epsffile{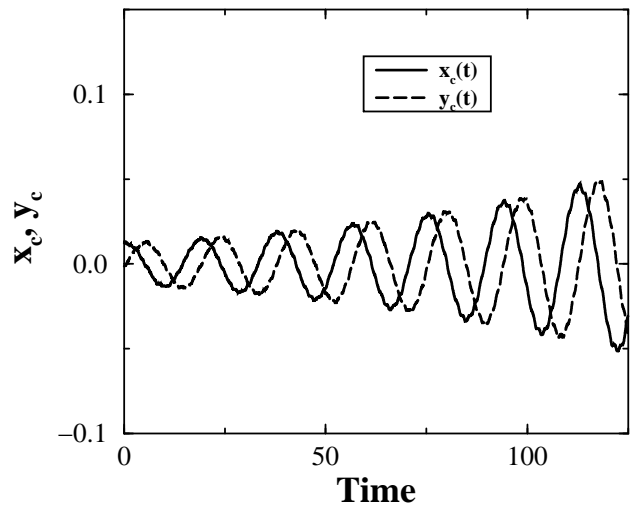}}
\caption{
Coordinates of spiral rotation center as functions of time,
in the regime of linear development of meandering instability.
Parameters of Barkley's model are $a = 0.66\,,
\,b = 0.01\,, \,\epsilon = 1/50\,$. System size is $10\times 10$,
number of grid-points is $81\times 81$. Time step $d t \approx 0.0016$.
\label{fig1}}
\end{figure}

In order to find the constant $\alpha$, we fitted the trajectory
of the spiral tip to the developing meander, in the form
$x(t) = x_0 + x_1 \sin(\omega_1 t + \phi_1) +
x_2 \exp(\alpha_1 t) \sin(\alpha_2 t + \phi_2)\,$,
where $x_0$ is the average position of the spiral tip,
$\omega_1$ is the main spiral rotation frequency, and $\alpha=\alpha_1+ 
i \alpha_2$. The same fitting was
carried out for $y(t)$. Thus, we have retrieved
the motion of the center of spiral rotational symmetry.
The coordinates of the center, $x_c$ and $y_c$, as functions of time
are given in Fig.~\ref{fig1}. We have found that, for the set of parameters
given in the caption to Fig.~\ref{fig1}, $\alpha_1 \approx 0.012\,$ and
$\alpha_2 \approx 0.336\,$. To find
the constant $\gamma$ we have applied a homogeneous advective field
to equation~(\ref{eq1}), along the $x$-axis, as we have done
in Ref.~\cite{MAK1}.
Then $\gamma$ is determined from fitting the spiral tip trajectory
to the meandering in the above form with an additional term
$\gamma_1 E t$ ($\gamma_2 E t$), in the expression for $x(t)$ ($\,y(t)\,$),
due to the drift of the spiral. 
The result for $\gamma$ is $\gamma_1 \approx 1.312\,$ and
$\gamma_2 \approx 0.154\,$.
In principle, it is easy to obtain the value of the parameter $\beta$
by matching the spiral center trajectory to the saturated meander.
However, we do not do it because it is not important for
the studied mechanism of the instability.

\begin{figure}[h]
\centerline{ \epsfxsize = 8.0cm \epsffile{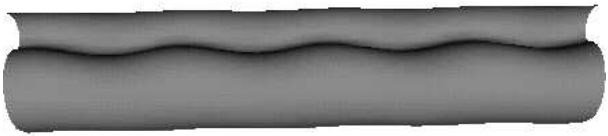}}
\caption{
Three-dimensional filament with the saturated 
instability. The system size is $10\times 10\times 40$,
number of grid-points is $81\times 81\times 320$.
Other parameters are the same as in Fig~\protect\ref{fig1}.
The initial perturbation is of the 3rd harmonics, with
the wavenumber $k \approx 0.47\,$.
\label{fig2}}
\end{figure}

\begin{figure}[h]
\hspace{0.2cm}
\rightline{ \epsfxsize = 9.0cm \epsffile{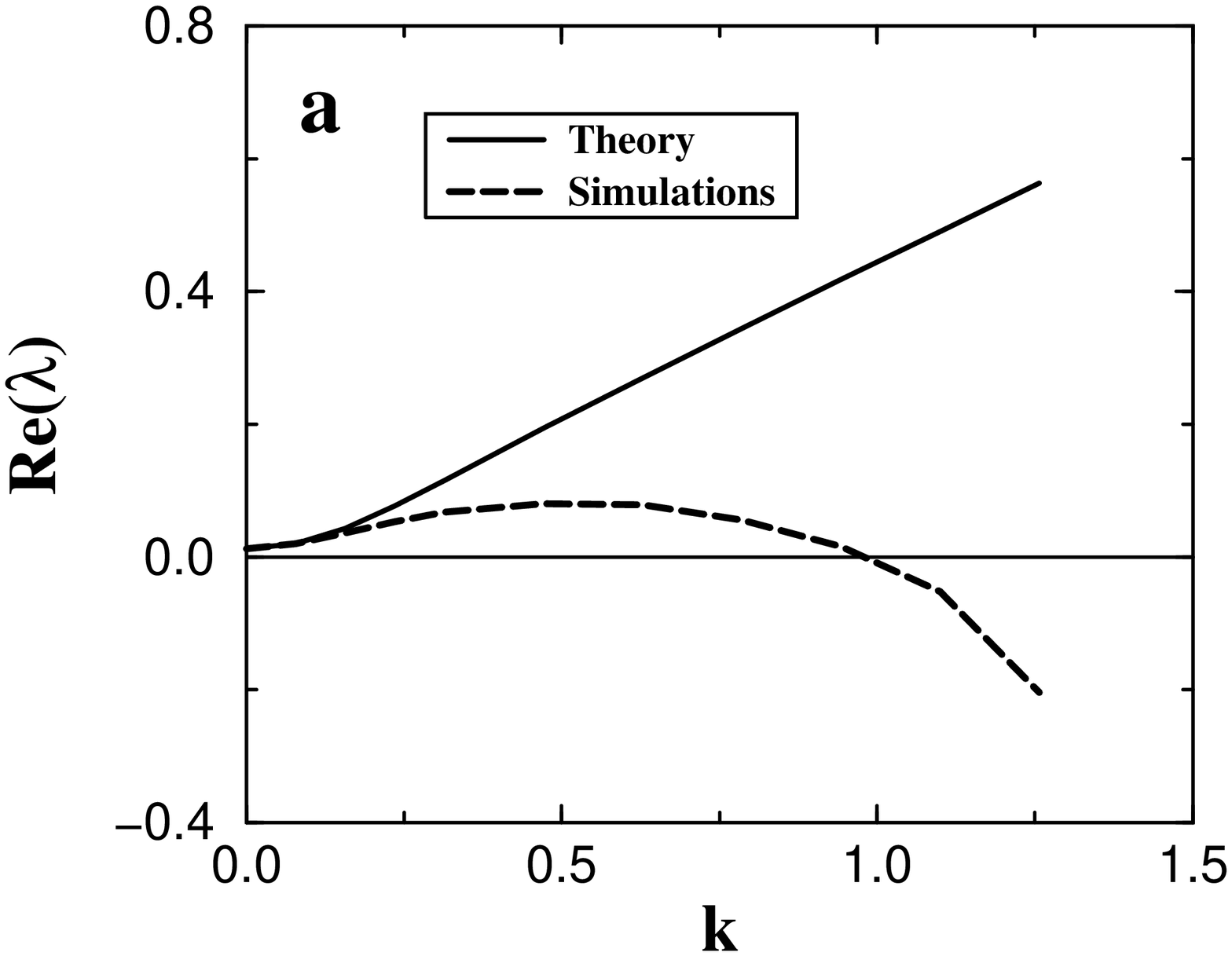}}

\vspace{-0.5cm}
\hspace{0.2cm}
\rightline{ \epsfxsize = 9.0cm \epsffile{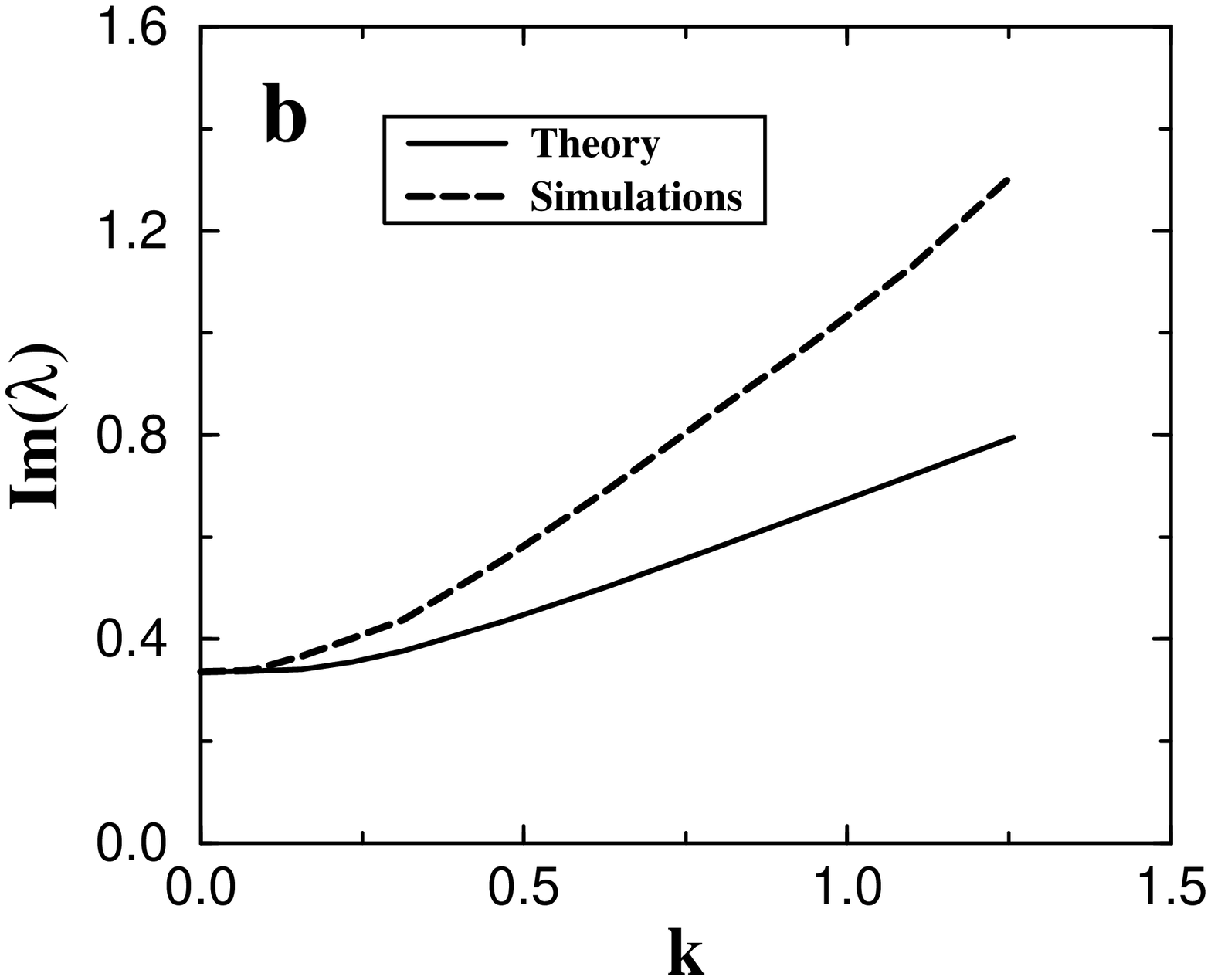}}
\caption{
The real (a) and imaginary (b) parts of
the eigenvalue $\lambda(k)$ of the three-dimensional instability
obtained analytically (solid line) and numerically (dashed line).
The system parameters are the same as in Fig~\protect\ref{fig2}.
\label{fig3}}
\end{figure}

We have performed numerical simulations of the three-dimensional
Eqs.~(\ref{eq1})--(\ref{eq2}). We have studied the behavior of an almost
straight scroll. We have prepared a two-dimensional purely
rotating (unstable)
spiral, with the parameter values in the meandering regime, using
the elaborated controlling technique described above, and 
translated it  along the third dimension to build an
initial scroll. 
Then, a periodic perturbation $\sim \exp [ i k z] $ has been applied. 
The three-dimensional plot of the helix, obtained as the result of
the instability saturation, is given in Fig.~\ref{fig2}.
In Fig.~\ref{fig3}
we have plotted the real and imaginary parts of the eigenvalue
$\lambda$ versus $k$, both for the theoretical prediction
given by~(\ref{eq6}), and for the results of the simulations.
As we have predicted, the observed instability of the filament
with initial perturbation of a finite $k$
appears to be substantially stronger than the
two-dimensional core meander instability ($\,k=0\,$).
We have also tried initial conditions other than purely periodic,
and have always found a saturated helix. Almost periodic
initial conditions have been mainly used to cut CPU time (the transient
from random initial conditions is very long).

As we can see from Fig.~\ref{fig3}, the $k \rightarrow 0$ asymptotics
of $Re[\lambda]$ and $Im[\lambda]$ obtained numerically are well matched 
by our theoretical prediction. As $k$ increases, however, the
small curvature approximation used for the derivation
of~(\ref{eq6}) no longer works, and the prediction
becomes invalid. This is why we see the divergence of the theoretical
curves from the numerical ones for larger $k$.

\begin{figure}[h]
\hspace{0.2cm}
\rightline{ \epsfxsize = 9.0cm \epsffile{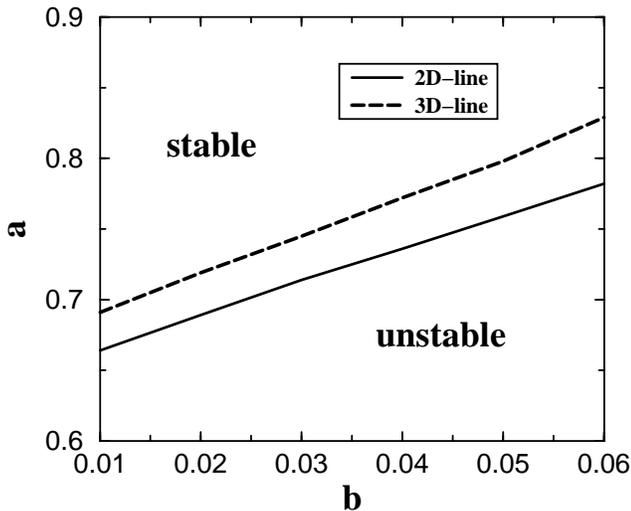}}
\caption{
Bifurcation lines of 2D (meander) instability (solid line)
and 3D instability (dashed line) in the plane $a$--$b$
of the Barkley's model parameters. The wavenumber $k\approx 0.63$ of
the initial perturbation corresponds to the 4th harmonics,
near the maximal growth rate of the instability, according to
Fig.~\protect\ref{fig3}(a). Other parameters
are the same as in Fig.~\protect\ref{fig3}.
\label{fig4}}
\end{figure}

We have also studied numerically the dynamics of filaments
in the space of parameters $a$ and $b$ of Barkley's model~\cite{barkley2}.
We have performed the stability analysis of the harmonic with
the growth rate near the maximal one (see Fig.~\ref{fig3}(a)\,).
In Fig.~\ref{fig4} we plot the bifurcation line of the obtained
three-dimensional instability together with that of the core meander
instability. We see that the curvature of the filament enhances
its instability, so that it is unstable even in a domain
of the $a$--$b$ plane, for which the two-dimensional spiral
is stable.

In conclusion, we have demonstrated that the
intrinsic three-dimensional instability of a straight scroll
leads to the formation of stable helicoidal structures.
We have shown that persistent curved vortex configurations are not necessarily
related to
a "negative line tension" of the filament, but originate from
the underdamped core dynamics.
Our current analysis is restricted to untwisted scrolls.
We expect that twisted scrolls are unstable even in a wider
space of the parameters,
and exhibit in general even more violent dynamics.
Helicoidal  structures with twist were observed recently in gel-immobilized
BZ reaction~\cite{pertsov1}.
The twist was generated
by an external temperature gradient along the scroll axis.
The helicoidal instability was observed above specific
value of the twist and disappeared when the temperature gradient
(and, therefore, the twist)
was removed. 
Presumably, by tuning the parameters of the  BZ reaction,
one may approach the limit
of the intrinsic three-dimensional instability for {\it untwisted} scrolls.
We can also speculate that the helicoidal instability of
scroll vortices in  large aspect ratio reaction-diffusion systems
is one of the mechanisms that supports more complex
vortex configurations and drives turbulent states.

We are grateful to R. Goldstein, A. Pertsov,
and S.-Y. Chen for illuminating discussions.
The  work of IA was supported by the U.S. Department
of Energy under contracts W-31-109-ENG-38  and
by the NSF, Office of STC
under contract No. DMR91-20000.


\begin{references}
\bibitem{comm} 
The  core instability in the complex Ginzburg-Landau
equation does not saturate to a stable meander~\cite{AKW}.
\end{references}
\end{document}